\documentclass[cits]{PoS}
\usepackage{amssymb}
\usepackage{graphicx}
\usepackage{latexsym}
\usepackage{amsmath}
\usepackage{bbm}
\usepackage{color}

\newcommand{\Z}{\mathbb{Z}}

\newcommand{\dd}{{\rm{d}}}

\newcommand{\tr}{{\rm Tr}}

\newcommand{\qhat}{\hat{q}}
\newcommand{\gE}{g_{\mbox{\tiny{E}}}}

\newcommand{\mE}{m_{\mbox{\tiny{E}}}}

\newcommand{\nb}{n_{\mbox{\tiny{B}}}}

\newcommand{\eq}{\begin{equation}}
\newcommand{\en}{\end{equation}}

\title{Momentum broadening of partons on the light cone from the lattice}

\ShortTitle{Momentum broadening of partons on the light cone from the lattice}

\author{\speaker{Marco Panero}\\
        Department of Physics and Helsinki Institute of Physics, University of Helsinki,\\
        P.O. Box 64, FI-00014 Helsinki, Finland\\
        E-mail: \email{marco.panero@helsinki.fi}}

\author{Kari Rummukainen\\
        Department of Physics and Helsinki Institute of Physics, University of Helsinki,\\
        P.O. Box 64, FI-00014 Helsinki, Finland\\
        E-mail: \email{kari.rummukainen@helsinki.fi}}

\author{Andreas Sch\"afer\\
        Institute for Theoretical Physics, University of Regensburg,\\
        D-93040 Regensburg, Germany\\
        E-mail: \email{andreas.schaefer@physik.uni-regensburg.de}}

\abstract{The jet quenching parameter describes the momentum broadening of a high-energy parton moving through the quark-gluon plasma. Following an approach originally proposed by Caron-Huot, we discuss how one can extract information on the collision kernel associated with the parton momentum broadening, from the analysis of certain gauge-invariant operators in dimensionally reduced effective theories, and present numerical results from a lattice study.
\vspace{4cm}
\begin{flushright}
HIP-2013-16/TH
\end{flushright}}

\FullConference{31st International Symposium on Lattice Field Theory LATTICE 2013\\
		 July 29 -- August 3, 2013\\
		 Mainz, Germany}

\begin{document}

\section{Introduction}
In this contribution, we present our recent study of the jet quenching phenomenon in the deconfined QCD plasma~\cite{Panero:2013pla}. More precisely, following an idea originally proposed in ref.~\cite{CaronHuot:2008ni}, we discuss the strategy for a lattice computation of the jet quenching parameter $\qhat$, which describes the average increase in squared transverse momentum per unit length experience by a high-energy quark propagating in the quark-gluon plasma (QGP). Recent related works include refs.~\cite{Ghiglieri:2013gia, Laine:2012ht, other_similar_qhat_computations}, while a different lattice approach to the same physical problem was put forward in ref.~\cite{Majumder:2012sh}.

Jet quenching is defined as the suppression of particles with high transverse momentum and of back-to-back correlations in heavy nuclei collisions~\cite{experiments}. It is considered as a ``gold-plated'' observable providing evidence for a strongly coupled quark-gluon plasma: the interpretation of the phenomenon is that a highly energetic parton moving through the deconfined medium undergoes energy loss and momentum broadening, due to interactions with the QGP constituents~\cite{Bjorken:1982qr}.

\section{Theoretical approach}
Jet quenching belongs to the class of \emph{hard probes} of the QGP: the hard scale involved is the momentum of the parton $Q$, which can be of the order of many GeV. For a certain collision of heavy nuclei, a complete theoretical description of the probability $\sigma_{(M + N \to \mbox{\tiny{hadron}})}$ of observing final-state hadrons with given momenta necessarily involves both perturbative and non-perturbative aspects. QCD factorization theorems, however, allow one to ``separate'' these different contributions in a well-defined way~\cite{CasalderreySolana:2007zz},
\begin{equation}
\sigma_{(M + N \to \mbox{\tiny{hadron}})}=
f_M(x_1,Q^2) \otimes f_N(x_2,Q^2) \otimes \sigma(x_1,x_2,Q^2) \otimes D_{\mbox{\tiny{parton}} \to \mbox{\tiny{hadron}}}(z,Q^2),
\end{equation}
where $f_{M}(x_1,Q^2)$ and $f_{N}(x_2,Q^2)$ are parton distribution functions, $D_{\mbox{\tiny{parton}} \to \mbox{\tiny{hadron}}}(z,Q^2)$ is a fragmentation function, while $\sigma(x_1,x_2,Q^2)$ describes the short-distance cross-section. In the following we concentrate our attention on the latter, focusing on the propagation of a highly energetic quark in the QGP. Following the standard formalism~\cite{Baier:1996kr}, one can describe the process in terms of multiple soft scatterings, in the eikonal approximation, assuming that the hard, ultrarelativistic parton moves on the light cone: then, the leading effect is the \emph{broadening} of the parton momentum. The average increase of the squared transverse momentum component per unit length defines the \emph{jet quenching parameter} $\qhat$, which can be evaluated if the \emph{collision kernel} $C(p_\perp)$, describing the differential collisional rate between the parton and plasma constituents, is known: 
\begin{equation}
\qhat = \frac{\langle p^2_\perp \rangle}{L} = \int \frac{ \dd^2 p_\perp}{(2\pi)^2} p^2_\perp C(p_\perp).
\end{equation}
The collision kernel is related to the two-point correlation function of light-cone Wilson lines. Therefore, as it is always the case for phenomena involving real-time dynamics, a direct computation of $C(p_\perp)$ via numerical simulations on a Euclidean lattice is not straightforward.

\section{Soft physics contribution from a Euclidean setup}
A full lattice computation of the collision kernel would necessarily require analytical continuation. However, we will now show that it is possible to extract the large, non-perturbative contributions to $C(p_\perp)$ \emph{directly} from Euclidean lattice simulations.

The idea can be understood in the limit of very high temperatures $T$, in which the physical coupling $g$ becomes small, and a hierarchy of parametrically separated momentum scales emerges: in addition to the typical scale of thermal excitations ($\pi T$), one can identify ``soft'' ($gT$) and ``ultrasoft'' ($g^2 T/\pi$) scales, which are relevant for the long-wavelength modes of the QGP. The ultrasoft scale, in particular, is intrinsically non-perturbative in nature: the long-wavelength modes of the plasma are strongly coupled at \emph{any} temperature, and large spatial Wilson loops are always confining. A well-defined, systematic way to treat the physics at these different scales is by means of dimensionally reduced effective theories~\cite{dim_red}. More precisely, the dynamics of the soft scale can be described by a static, purely Euclidean, effective theory, which corresponds to three-dimensional Yang-Mills theory coupled to an adjoint scalar field (``electrostatic QCD'' or EQCD). In turn, the low-energy limit of EQCD is captured by an effective theory which is just three-dimensional Yang-Mills theory, encoding the physics at the ultrasoft scale (``magnetostatic QCD'' or MQCD). The typically large next-to-leading order (NLO) corrections affecting perturbative expansions in thermal QCD are related to soft, essentially classical, fields.

As observed in ref.~\cite{CaronHuot:2008ni}, for physics on the light cone, it turns out that the contributions from soft modes are insensitive to the precise value of the parton velocity: in particular, they would remain unchanged even for superluminal partons. This idealized case is obviously unphysical, but it would make the parton world-line space-like: then, it ought to be possible to extract the soft contribution to the correlation function of light cone Wilson lines from a Euclidean setup. Indeed, this can be proven rigorously~\cite{CaronHuot:2008ni, Ghiglieri:2013gia}: in momentum space, the two-point correlation function of spatially separated ($|t|<|z|$) light-like Wilson lines $G^<(t,x_\perp,z)$ can be written in terms of the difference between a retarded ($\tilde{G}_{\mbox{\tiny{R}}}$) and an advanced ($\tilde{G}_{\mbox{\tiny{A}}}$) correlator,
\begin{equation}
\label{Minkowski_to_Euclidean_derivation}
G^<(t,x_\perp,z) = \int \!\! \dd \omega \dd^2 p_\perp \dd p^z \left[ \frac{1}{2} + \nb (\omega) \right] \left[ \tilde{G}_{\mbox{\tiny{R}}}(\omega,p_\perp,p^z) - \tilde{G}_{\mbox{\tiny{A}}}(\omega,p_\perp,p^z) \right] e^{-i \left( \omega t - x_\perp \cdot p_\perp - z p^z \right)},
\end{equation}
where $\nb$ denotes the Bose-Einstein distribution. Shifting the momentum component along the direction of motion as $p^{\prime z}=p^z - \omega t/z$, the integration  over frequencies can be carried out by analytical continuation into the upper (lower) complex half-plane for the retarded (advanced) contribution. This leads to a sum over Matsubara frequencies involving the Euclidean correlator $\tilde{G}_{\mbox{\tiny{E}}}$,
\begin{equation}
\label{Minkowski_to_Euclidean_derivation_2}
G^<(t,x_\perp,z) = T \sum_{n \in \Z } \int \dd^2 p_\perp \dd p^{\prime z} \tilde{G}_{\mbox{\tiny{E}}}(2 \pi n T,p_\perp,p^{\prime z} + 2 \pi i n T t /z ) e^{i \left( x_\perp \cdot p_\perp + z p^{\prime z} \right)}.
\end{equation}
All non-zero modes are suppressed at large separations. The soft contribution, however, is entirely encoded in the $n=0$ mode, which is time-independent and therefore can be evaluated in EQCD.

\section{Lattice implementation}
The lattice computation of the two-point correlation function of light-cone Wilson loops is then straightforward. We regularize the super-renormalizable continuum EQCD Lagrangian
\begin{equation}
\mathcal{L} = \frac{1}{4} F_{ij}^a F_{ij}^a + \tr \left( (D_i A_0)^2 \right) + \mE^2 \tr \left( A_0^2 \right) + \lambda_3 \left( \tr \left( A_0^2 \right) \right)^2
\end{equation}
on the lattice using the Wilson formulation, and fix its parameters (the 3D gauge coupling and the mass- and quartic-term coefficients) by \emph{matching} to high-temperature QCD~\cite{dim_red}. We chose a setup corresponding to QCD with $n_f=2$ light dynamical quarks, at two temperatures $T \simeq 398$~MeV and $2$~GeV, respectively equal to about twice and ten times the deconfinement temperature~\cite{Hietanen:2008tv}.

\begin{figure}[-t]
\centerline{\includegraphics[height=0.25\textheight]{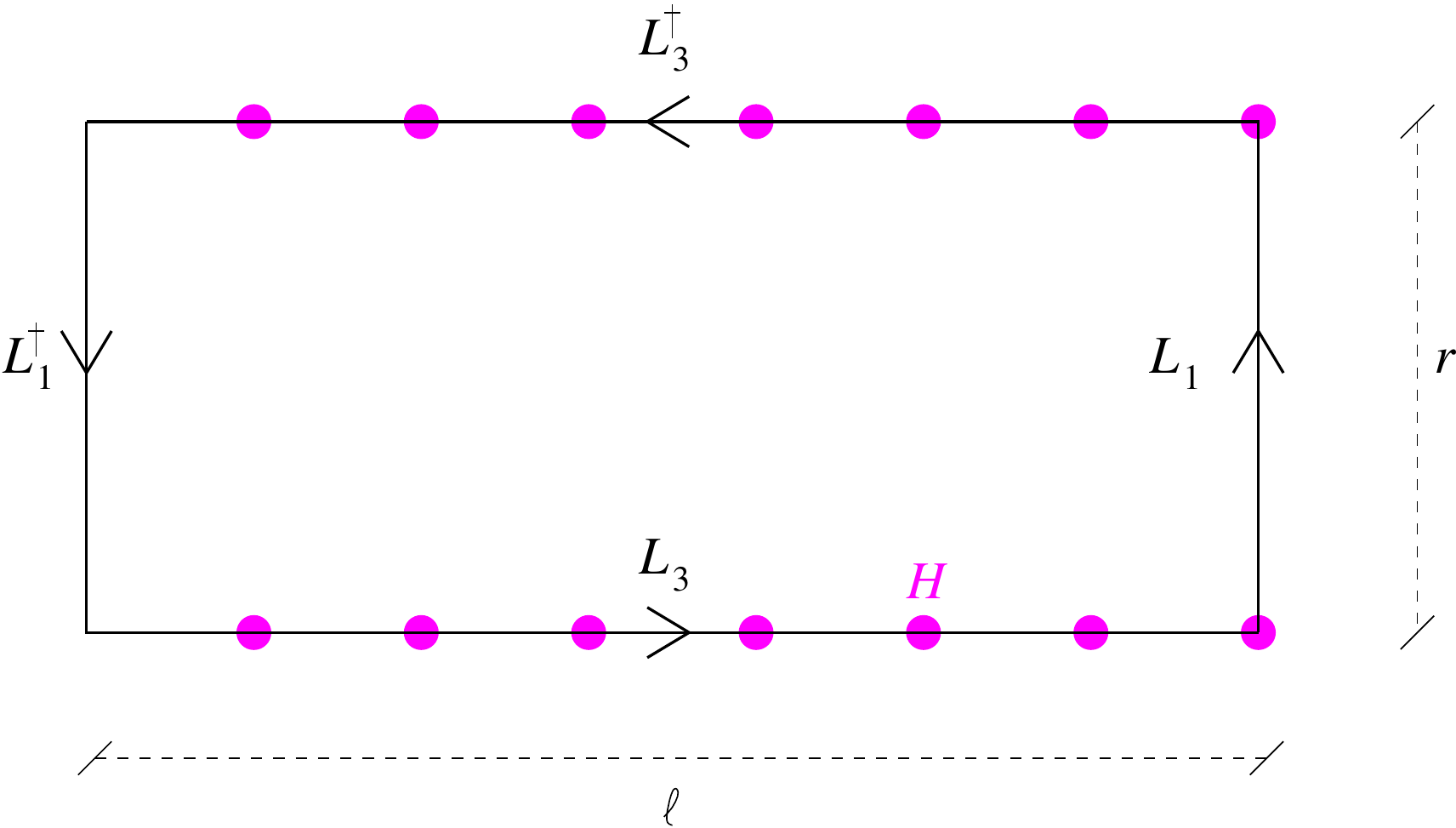}}
\caption{The gauge-invariant operator describing the two-point correlation function of light-cone Wilson lines appearing in eq.~(\protect\ref{decorated_loop}) involves a set of Hermitian parallel transporters along the real time direction $H(x)$, obtained from exponentiation of the scalar field $A_0(x)$.}
\label{fig:simplified_decorated_loop}
\end{figure}

The two-point correlator of light-cone Wilson lines can be made gauge-invariant by ``closing it'' (with the inclusion of transverse parallel transporters) to a loop and taking its trace, $W(\ell,r)$. Then, its lattice implementation becomes straightforward. An important point to remark, however, is the following: although the effective theory that we consider is purely spatial, the operator must describe \emph{real time} evolution. For this reason, the lattice discretization of the two light-cone Wilson lines involves insertions of the local Hermitian operator $H(x) = \exp [- a \gE^2 A_0(x) ]$, representing a parallel transporter along the real-time direction:
\begin{equation}
\label{decorated_loop}
\left\langle W(\ell, r) \right\rangle = \left\langle \tr \left( L_3 L_1 L^\dagger_3 L^\dagger_1 \right) \right\rangle ,
\end{equation}
with $L_3 = \prod (U_3 H)$ and $L_1 = \prod U_1$, see fig.~\ref{fig:simplified_decorated_loop}. This lattice operator has well-defined renormalization properties~\cite{D'Onofrio_et_al}. From the expectation value of $W(\ell, r)$ (that we compute with the multilevel algorithm~\cite{Luscher:2001up}) one can extract a ``potential'' $V(r)$,
\begin{equation}
\label{potential}
\left\langle W(\ell, r) \right\rangle = \exp \left[ - \ell V(r) \right],
\end{equation}
which is nothing but the transverse Fourier transform of the collision kernel $C(p_\perp)$.

\section{Results and discussion}
\begin{figure}[-t]
\centerline{\includegraphics[height=0.27\textheight]{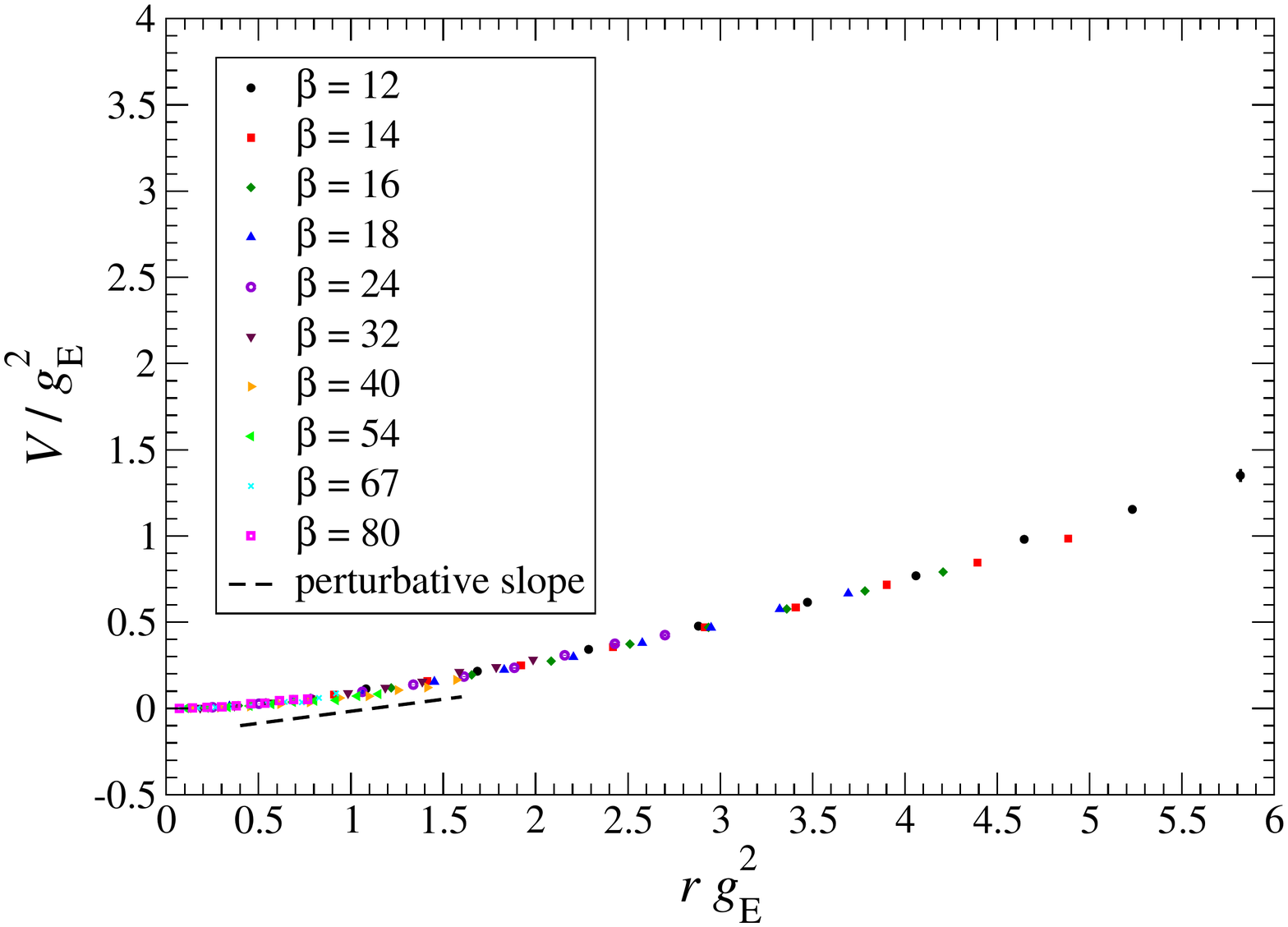} \hfill \includegraphics[height=0.27\textheight]{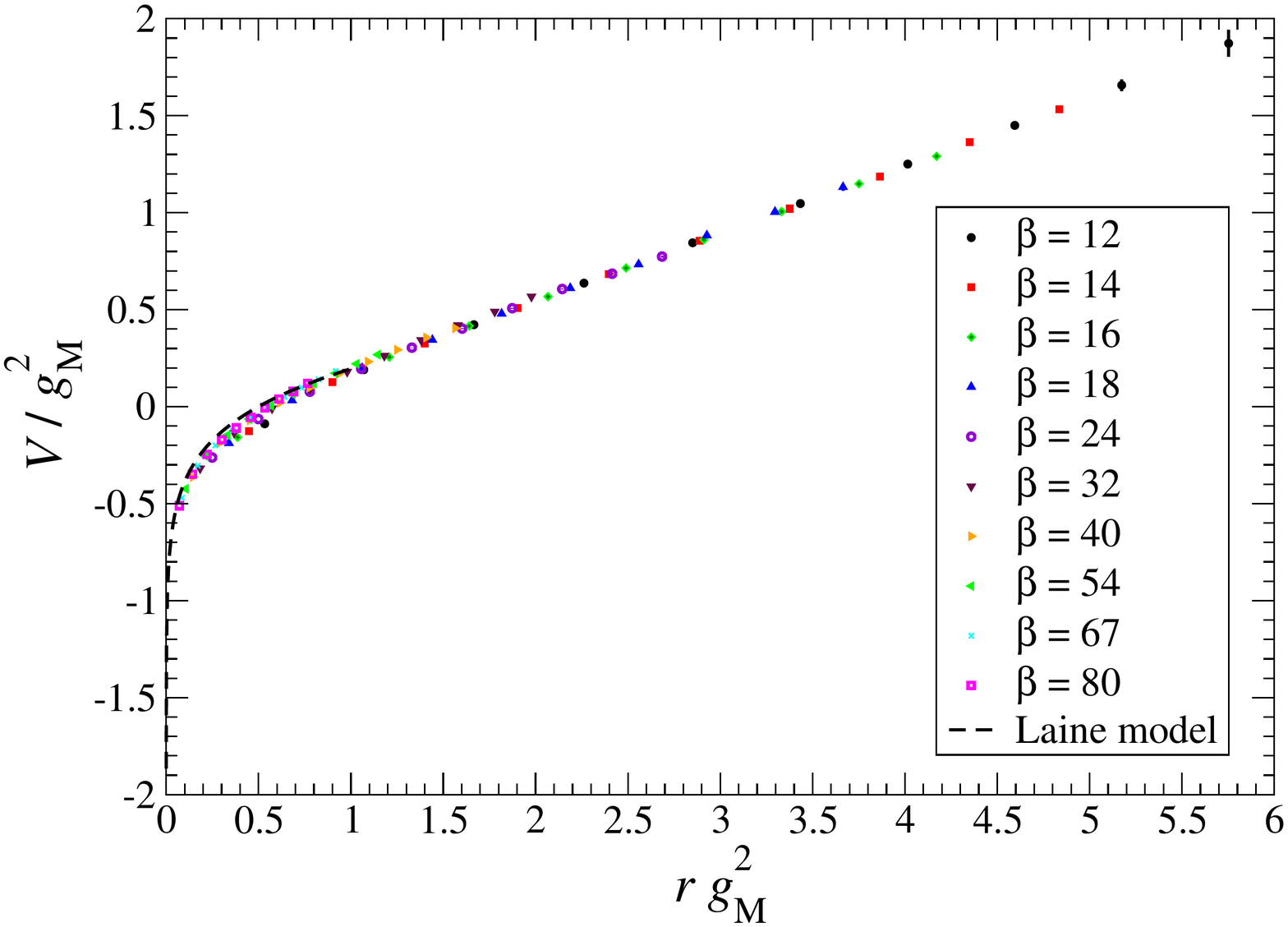}}
\caption{Left-hand-side panel: Our results for $V(r)$ extracted from eq.~(\protect\ref{potential}), for our lower temperature $T \simeq 398$~MeV. The slope predicted perturbatively for $r \gE^2 \simeq 1$ is also shown. Right-hand-side panel: The static-quark potential extracted in three-dimensional Yang-Mills theory. The dashed curve is a short-distance parametrization based on the analysis carried out in ref.~\cite{Laine:2012ht}.}
\label{fig:V_r}
\end{figure}

Our results for $V(r)$ are shown in the left-hand-side panel of fig.~\ref{fig:V_r}: they exhibit very good scaling properties and are compatible with the perturbative predictions that $V(r)$ should vanish for $r=0$, and that at distances $r \simeq 1/\gE^2$ it should include a linear contribution, with the slope of the dashed line displayed in the figure. Note that this short-distance behavior is dramatically different from the one of the usual potential extracted from ordinary Wilson loops in 3D Yang-Mills theory (which we also computed, for comparison with the MQCD study discussed in ref.~\cite{Laine:2012ht}), shown in the right-hand-side panel of fig.~\ref{fig:V_r}.

Since $\qhat$ is given by the second moment associated with the $C(p_\perp)$ kernel, it can be related to the \emph{curvature} of $V(r)$. Following the approach discussed in ref.~\cite{Laine:2012ht}, we arrive at the result that the soft NLO contribution to $\qhat$ is quite large: $0.55(5)\gE^6$ for $T \simeq 398$~MeV, and $0.45(5)\gE^6$ for $T \simeq 2$~GeV. Taking also the other known contributions into account, our analysis leads to a rough estimate for the final value of $\qhat$ in the ballpark of $6$~GeV$^2$/fm for RHIC temperatures. Interestingly, this result is comparable with strong-coupling predictions from holography~\cite{Liu:2006ug}, as well as with phenomenological computations~\cite{Dainese_Eskola}.

In summary, we have shown that a lattice approach is possible for certain real-time problems involving physics on the light cone. In particular, here we have focused on soft physics in thermal QCD (but related ideas have also been discussed in the context of QCD at zero temperature~\cite{Ji_Lin}). We emphasize that the outlined approach is \emph{systematic}: this is not a model. The effective theory study that we carried out in this work is based on the modern approach to thermal QCD, providing a consistent framework to describe the physics at different momentum scales relevant for the QGP. Although, by construction, the effective theory that we simulated does not capture phenomena at the hard thermal scale, we stress that this is not a limit, but a virtue, of the formalism, allowing one to disentangle effects of different physical origin.

As for possible extensions of this work, we plan to improve our extrapolation to the continuum limit at short distances by simulations on finer lattices, and/or using improved actions (as originally suggested in ref.~\cite{Mykkanen:2012dv}). We also plan to study in detail the temperature dependence of $\qhat$ and compare it with holographic expectations~\cite{Liu:2006ug}. In the same spirit, it would also be interesting to repeat this study in theories with more than $N=3$ colors, in which the dynamics simplifies considerably~\cite{large_N_general}. Recent lattice studies show that the static equilibrium properties of the QGP have very little dependence on $N$, both in four~\cite{large_N_finite_T_4D} and in three spacetime dimensions~\cite{large_N_finite_T_3D}, and, given that all holographic computations are carried out in the large-$N$ limit, it would be interesting to see if this is also the case for quantities related to real-time dynamics.

\vskip1.0cm 
\noindent{\bf Acknowledgements}\\
This work is supported by the Academy of Finland (project 1134018) and by the German DFG (SFB/TR 55). Part of the numerical simulations was carried out at the Finnish IT Center for Science (CSC) in Espoo, Finland.

\end{document}